\documentclass[twocolumn,showpacs,amsmath,amssymb]{revtex4}

    \usepackage{graphicx}
    \usepackage{dcolumn}
    \usepackage{bm}
    \newcommand{\VEC}[1]{\mbox{\boldmath${#1}$}}
    \newcommand{\mix}{\theta}

    \newcommand{\IgnoreThis}[1]{#1}

\begin{document}

\title{Unequal intensity splitting can reduce back action in
interferometers}

\author{Ole Steuernagel}

\address{School of Physics, Astronomy and Mathematics, University of
Hertfordshire, Hatfield, AL10 9AB, UK}

\email{O.Steuernagel@herts.ac.uk}

\date{\today}

\begin{abstract}
Typical two-path interferometers are intensity-balanced because this
maximizes the visibility of their interference patterns. Unbalancing
the interferometer can be advised when back action on the object
whose position is monitored is to be reduced. Variations of the
intensity splitting ratios in two-path interferometers are analyzed
in order to determine optimal interferometric performance while
minimizing back action: it turns out that homodyning-like schemes
perform best.
\end{abstract}


\maketitle

\IgnoreThis{\section{Introduction}}

Typical two-path interferometers are balanced: the intensity in
probe and reference arm are the same. This maximizes the visibility
of the interference pattern because only for balanced illumination
complete destructive interference is observed~\cite{Wootters_PRD79}.
Also, this tends to simplify the setup, a single balanced beam
splitter to split and reunite the light paths can be used, such as
in the conventional Michelson-Morley setup. Since recent
technological progress allows us to monitor very sensitive systems,
e.g. nano-mechanical oscillators integrated into optical
setups~\cite{Kippenberg_SCI08,Marquardt_Physics.2.40}, increasingly,
the back action~\cite{BraginskyKhalili.book} on the object whose
characteristics are monitored has to be taken into account. Here we
will only consider probing of mirror positions through elastic
scattering, yet, even in ideal setups radiation pressure back action
cannot be avoided~\cite{BraginskyKhalili.book,Law_PRA95}.

Intuitively, it is clear that unbalancing an interferometer: sending
less light towards the probed mirror, will reduce the interference
fringe visibility and reduce the back action of the light onto the
mirror (only intensity-unbalancing is considered here, not using
unequal path lengths). Whereas the first effect is unwelcome, the
second maybe desired and we therefore want to consider the tradeoff
between best visibility and least back action on the object. It
should, perhaps, be emphasized that these considerations do not
apply to gravitational wave detection in that the quadripolar nature
of gravity waves requires probing in \emph{both} arms of the
interferometer~\cite{WallsMilburn.book,ScullyZubairy.book}.

It is known that the visibility of the interference pattern remains
high when moderate power splitting imbalances are used in an
interferometer~\cite{Wootters_PRD79}. It is therefore plausible that
an imbalance will allow us to reduce the back action of the light
onto an object while retaining some visibility. To quantify the back
action we will consider radiation pressure, proportional to the
light's intensity~\cite{Law_PRA95}, and its fluctuations. These
intensity fluctuations tend to randomize the mirror's position and
momentum distributions.

We will treat the light fields quantum-optically but only consider
coherent light fields because this yields our considerations simple,
and, yet, relevant and general. `Simplicity' of calculations arises
from the fact that coherent states are quasi-classical, `relevance'
is due to the fact that coherent states are the most important light
states used in interferometry.  Partly this is due to the fact that
other light states are hard to synthesize but also because they are
so fragile (\cite{Steuernagel_Scheel_JOB04,Dorner_PRL09} and
references therein) that only balanced setups perform well, see
e.g.~\cite{Steuernagel_Scheel_JOB04}.

Our treatment is also `general' since coherent states are fluctuation
minimized (the cycle-averaged fluctuation powers of any
noise-minimized monochromatic light fields are the same as those of
coherent
states~\cite{WallsMilburn.book,ScullyZubairy.book,GerryKnight.book}).
Our analysis can therefore be carried over to non-classical light
states such as squeezed and squeezed-coherent states. Such
non-classical states may allow for greater interferometric resolution
power than coherent states but their cycle-averaged fluctuations are
the same and it is therefore straightforward to adopt our discussions
accordingly~\cite{WallsMilburn.book,ScullyZubairy.book,GerryKnight.book}.

\begin{figure}[t]
\centering
\includegraphics[width=0.48\textwidth,height=0.125\textwidth]{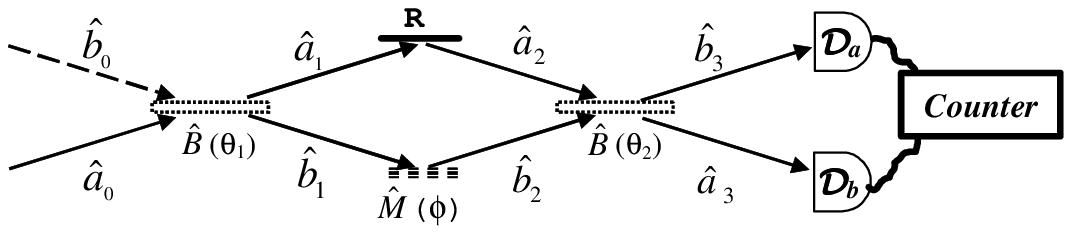}
\caption{An unbalanced interferometer is illuminated with light in a
  Glauber-coherent state in port $\hat a_0$ whereas port $\hat b_0$ is in
  vacuum. The first beam splitter ${\VEC{\hat B}}$ mixes the modes
  with mixing angle~$\mix_1$. This is followed by a fixed reflecting
  mirror~$R$ in the upper path and a sensitive moving mirror~$\hat M$
  in the lower path. Movement of~$M$ can shift the phase~$\phi$
  of the reflected light. Finally, the two modes are mixed again at
  the second beam splitter and fed into the detectors~$\cal D$.}
\label{fig_Setup}
\end{figure}

\IgnoreThis{\section{Resolution power of unbalanced
interferometers}}

To describe the light modes traversing the interferometer we use
creation~${\hat a}^\dagger / {\hat b}^\dagger$ and annihilation~${\hat
  a} / {\hat b}$ operators for the upper / lower paths,
see~Fig.~\ref{fig_Setup}, which obey bosonic commutation relations,
e.g.~$ [ {\hat a},{\hat a}^\dagger ] = {\hat 1} $. An unbalanced
beam splitter can be described by the unitary matrix,~${\VEC{\hat
B}}$, parameterized by the mixing angle~$\mix$ which mixes the light
modes according to
\begin{equation}\label{eq_BeamSplitter}
    \VEC{\hat B}(\mix)=\left(%
\begin{array}{cc}
  \cos(\mix) & \sin(\mix) \\
  -\sin(\mix) & \cos(\mix) \\
\end{array}%
\right) .
\end{equation}

For an unbalanced interferometer with only coherent light input we
can, without loss of generality, assume that only one input
channel~($\hat a_0$) is used, see~Fig.~\ref{fig_Setup}. The input
light state is a product of a coherent state with amplitude
`$\alpha$' in port~$\hat a_0$ and vacuum in port~$\hat b_0$, namely
$\psi_0=|\alpha, 0 \rangle$. The light is redistributed amongst both
interferometer paths by the first beam splitter according to
\begin{eqnarray}\label{eq_FirstTrafo}
    \left( \begin{array}{c} \hat a_1 \\ \hat b_1 \\ \end{array} \right)
= \VEC{\hat B}(\mix_1) \left( \begin{array}{c} \hat a_0 \\ \hat b_0
\\ \end{array} \right) .
\end{eqnarray}
This implies that the probe arm light intensity is given by $I_{b_1}
= \langle \hat b_1^\dagger \hat b_1 \rangle =|\alpha|^2
\sin(\mix_1)^2 $ and the standard deviation of its fluctuations is
$\Delta I_{b_1} = \sqrt{\langle (\hat b_1^\dagger \hat b_1)^2  -
I_{b_1}^2 \rangle} = | \alpha \sin(\mix_1) |$.

The action of the probed mirror is described by the transformation
\begin{equation}\label{eq_M_mirror}
    \VEC{\hat M}(\phi)=\left(%
\begin{array}{cc}
  e^{-i \phi} & 0 \\
  0 & 1 \\
\end{array}%
\right) \, ,
\end{equation}
where $\phi$ parameterizes the phase delay due to path length
variations when the mirror $M$ is moved, see~Fig.~\ref{fig_Setup}.
The reference arm~$R$ is assumed fixed and in~$\VEC{\hat M}(\phi)$
thus described by multiplication with unity. After superposition of
the interferometer modes at the second beam splitter the entire
setup, displayed in Fig.~\ref{fig_Setup}, gives rise to the mode
transformations
\begin{eqnarray}\label{eq_FullTrafo}
    \left( \begin{array}{c} \hat a_3 \\ \hat b_3 \\ \end{array} \right)
= \VEC{\hat B}(\mix_2) \VEC{\hat M}(\phi) \VEC{\hat B}(\mix_1)
\left( \begin{array}{c} \hat a_0 \\ \hat b_0 \\ \end{array} \right)
.
\end{eqnarray}

The output modes expressed in terms of the input modes then have the
form
\begin{eqnarray}
     \hat a_3 = & \! \! \!
\cos(\mix_2) \left(\cos(\mix_1){\hat a_0}+\sin(\mix_1){\hat
b_0}\right)+ \nonumber
\\
& e^{-i\phi} \sin(\mix_2) \left(-\sin(\mix_1){\hat
a_0}+\cos(\mix_1){\hat b_0}\right)  , \label{eq_FullTrafo_a3}
\end{eqnarray}
and
\begin{eqnarray}
     \hat b_3 = & \! \! \!
-\sin(\mix_2) \left(\cos(\mix_1){\hat a_0}+\sin(\mix_1){\hat
b_0}\right)+ \nonumber
\\
& e^{-i\phi} \cos(\mix_2) \left(-\sin(\mix_1){\hat
a_0}+\cos(\mix_1){\hat b_0}\right)  . \label{eq_FullTrafo_b3}
\end{eqnarray}

To extract the maximal interferometric signal we form the difference
of the output intensities detected by the detectors~${\cal D}_b$
and~${\cal D}_a$, that is, we consider the
observable~\cite{GerryKnight.book}~$\hat {\cal O} = \hat b_3^\dagger
\hat b_3 - \hat a_3^\dagger \hat a_3 $. In our case its expectation
value has the form
\begin{align}
{\cal O} = & \; \alpha^2
\left[4\sin(\mix_2)\cos(\mix_1)\cos(\mix_2)\sin(\mix_1)\cos(\phi)-1
\right. \nonumber \\
&\left.+2\cos(\mix_1)^2-4\cos(\mix_2)^2\cos(\mix_1)^2+2\cos(\mix_2)^2\right],
\label{eq_Expec_O}
\end{align}
and its standard deviation is~$\Delta {\cal O} = |\alpha|$. Since
the ratio of standard deviation and maximal signal gradient yields
the angular resolution~$\Delta \phi$~\cite{GerryKnight.book}, we
find
\begin{eqnarray}
\Delta \phi = \Delta {\cal O}/ \left|\frac{\partial {\cal
O}}{\partial \phi}\right| = \frac{1}{|\alpha
\sin(2\mix_1)\sin(2\mix_2)\sin(\phi)|}. \label{eq_Delta_Phi}
\end{eqnarray}

This confirms that best interferometric resolution is achieved for a
balanced interferometer, $\mix_1=\mix_2=\pi/4$, where the optimal
working point is~$\phi^\star = \pi/2$ and large intensities are
desired to minimize $\Delta \phi^\star=
1/|\alpha|$~\cite{GerryKnight.book}.

\IgnoreThis{\section{Resolution power traded off against back
action}}

When we consider the intensity in the probe arm as the back action
quantity we want to minimize, while maximizing the angular
resolution power~$1/\Delta \phi$, we are led to consider the
\emph{intensity based} performance ratio
\begin{eqnarray}
\rho_I = \frac{1}{ \Delta \phi \, I_{b_1}} = \frac{|
\sin(2\mix_1)\sin(2\mix_2)\sin(\phi)|}{|\alpha \, \sin(\mix_1)^2|}.
\label{eq_Ratio_Intensity}
\end{eqnarray}
For a balanced interferometer $\mix_1 = \mix_2 = \pi/4$ operating at
the working point~$\phi=\phi^\star=\pi/2$ we find
$\rho_{I}=2/|\alpha|$.

Clearly the operation of an interferometer as a balanced
interferometer is not optimal when it comes to avoiding back action.
The intensity based performance ratio~$\rho_I$ becomes maximized
when we choose a balanced second beam mixer~$\mix_2=\pi/4$ and a
nearly transparent first beam splitter~$\mix_1 \sim 0$. In this
case~$\rho_I$ becomes formally unbound, indicating better
performance. Note, however, that reducing the laser power to zero
also has the effect of formally increasing~$\rho_I$ beyond any
bound. This leads us to conclude that~$\rho_I$ might not be the most
suitable measure for the quantification of interferometric
performance, but it indicates that the trend towards better
performance is given by an imbalance that reduces the flow of light
towards the probed object.

We now turn to a discussion of the \emph{fluctuation based}
performance ratio
\begin{eqnarray}
\rho_{\Delta I} = \frac{1}{ \Delta \phi \, \Delta I_{b_1}} = \frac{|
\sin(2\mix_1)\sin(2\mix_2)\sin(\phi)|}{|\sin(\mix_1)|}.
\label{eq_Ratio_IntensityFluc}
\end{eqnarray}
In the balanced interferometer case this yields $\rho_{\Delta I} =
\sqrt{2}$. For arbitrary mixing angles but assuming that the same
beam splitter is used twice (e.g. in a Michelson-Morley setup)
$\mix_1=\mix_2=\arctan(1/\sqrt{2})\approx 35.3^\circ$ yields the
best result. In this case $\rho_{\Delta I}= \sqrt{3}
 \sin(2\arctan(1/\sqrt{2})) \approx 1.54 \approx 1.09 \sqrt{2}$,
where the last expression shows a nine percent increase in
performance over the balanced interferometer case.

Best performance is achieved for an interferometer in which the beam
mixer is balanced,~$\mix_2=\pi/4$, and the beam splitter is nearly
transparent, $\mix_1\approx 0$. In this case an expansion in
$\mix_1$ yields~$\rho_{\Delta I} \approx
(2-\mix_1^2)\sin(2\mix_2)\sin(\phi)$. At the working point
$\phi^\star$ performance can reach~$\rho_{\Delta I} = 2$ which is by
$\sqrt{2}$ better than the balanced case.

\begin{figure}[t]
\centering
\includegraphics[width=0.48\textwidth,height=0.25\textwidth]{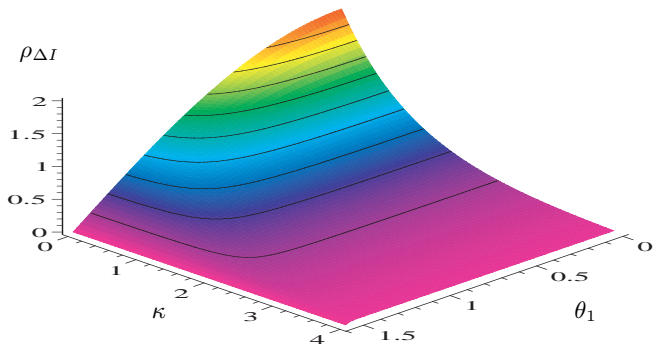}
\put(-240,109){\rotatebox{0}{\mbox{$\rho_{\Delta I}$}}}
\put(-190,10){$\kappa$} \put(-30,10){$\mix_1$}
\caption{(color online) Plot of the {fluctuation based} performance
ratio~$\rho_{\Delta I}$, see eq.~\ref{eq_Ratio_IntensityFluc}, at
the working point~$\phi^\star=\pi/2$ with a balanced beam
mixer~$\mix_2=\pi/4$. $\rho_{\Delta I}$ is plotted as a function of
losses~$e^{-\kappa}$ in the probe arm and mixing angle~$\mix_1$ of
the first beam splitter.}
\label{fig_ResolutionPowerRatio_losses}
\end{figure}

\IgnoreThis{\section{Losses in the interferometer and imperfect
detectors}}

If losses occur at the moving mirror~$M$ because radiation is
absorbed or scattered into other modes this would have to be
included through the mixing in of additional vacuum modes. In the
case of coherent states the losses and associated mixing in of
vacuum modes can be modelled by an amplitude attenuation
factor~$e^{-\kappa}$ with a positive $\kappa$. This means we can
simply substitute the phase shifting
transformation~(\ref{eq_M_mirror}) by
\begin{equation}\label{eq_M_mirror_kappa}
    \VEC{\hat M}(\phi)=\left(%
\begin{array}{cc}
  e^{-i \phi - \kappa} & 0 \\
  0 & 1 \\
\end{array}%
\right) \, .
\end{equation}
Clearly this brakes unitarity of the state evolution and cannot be
applied to states other than coherent states. In this case the
expression for the angular resolution~$\Delta \phi$, compare
eq.~(\ref{eq_Delta_Phi}), becomes
\begin{eqnarray}
\Delta \phi (\kappa) = \frac{\sqrt{\cos(2\mix_1) (e^{2\kappa}-1) +
e^{2\kappa}+1}}{|\sqrt{2} \alpha
\sin(2\mix_1)\sin(2\mix_2)\sin(\phi)|}. \label{eq_Delta_Phi_kappa}
\end{eqnarray}
At the working point $\phi^\star$ and for a balanced beam
mixer~$\mix_2=\pi/4$ this yields the fluctuation based performance
ratio
\begin{eqnarray}
\rho_{\Delta I}  = \frac{\sqrt{2}| \sin(2\mix_1) | }{
|\sin(\mix_1)\sqrt{\cos(2\mix_1) (e^{2\kappa}-1) + e^{2\kappa}+1}|},
\label{eq_Ratio_IntensityFluc_kappa}
\end{eqnarray}
which is plotted in Fig.~\ref{fig_ResolutionPowerRatio_losses}. Note
that this expression for~$\rho_{\Delta I}$ was derived assuming that
the back action expression~$\Delta I_{b_1}$ is unchanged although it
may actually depend on whether light was absorbed or scattered, and
how it was scattered.

If the detectors are (both equally) inefficient, i.e. only a
fraction~$\eta$ of the light gets detected, our results still change
little. For Glauber-coherent light the performance
ratio~$\rho_{\Delta I}$ simply becomes~$\eta\cdot\rho_{\Delta I}$.

\hspace{1cm}

\IgnoreThis{\section{Conclusions}}

Two-path interferometers with one movable mirror, fed with Glauber
coherent light, are analyzed. The trade-off between resolution power
of and the back action onto the probed mirror is analyzed. Typically
best performance is attained when light is directed away from the
probed object into the reference arm of the interferometer. If the
setup uses the same beam splitter to split and merge the light beams
a mixing angle of~$\mix=\arctan(1/\sqrt{2})\approx 35.3^\circ$ gives
best results, compare Fig.~\ref{fig_Setup}.

In general, large imbalances, using a \emph{homodyne-like setup}
(weak interferometric signal superposed with strong local
oscillator), yield best performance. Setups that send small amounts
of light towards the probed object perform nearly equally well: the
area for small values of~$\mix_1$ shows very weak gradients
in~$\mix_1$ (see Fig.~\ref{fig_ResolutionPowerRatio_losses}).
Therefore some freedom remains to decide on just how little light
one wants to send towards the object and how poor a phase resolution
one can tolerate.

\bibliography{UnbalancedInterferometerBib}

\begin{thebibliography}{10}
\expandafter\ifx\csname natexlab\endcsname\relax\def\natexlab#1{#1}\fi
\expandafter\ifx\csname bibnamefont\endcsname\relax
  \def\bibnamefont#1{#1}\fi
\expandafter\ifx\csname bibfnamefont\endcsname\relax
  \def\bibfnamefont#1{#1}\fi
\expandafter\ifx\csname citenamefont\endcsname\relax
  \def\citenamefont#1{#1}\fi
\expandafter\ifx\csname url\endcsname\relax
  \def\url#1{\texttt{#1}}\fi
\expandafter\ifx\csname urlprefix\endcsname\relax\def\urlprefix{URL }\fi
\providecommand{\bibinfo}[2]{#2}
\providecommand{\eprint}[2][]{\url{#2}}

\bibitem[{\citenamefont{Wootters and Zurek}(1979)}]{Wootters_PRD79}
\bibinfo{author}{\bibfnamefont{W.~K.} \bibnamefont{Wootters}} \bibnamefont{and}
  \bibinfo{author}{\bibfnamefont{W.~H.} \bibnamefont{Zurek}},
  \bibinfo{journal}{Phys. Rev. D} \textbf{\bibinfo{volume}{19}},
  \bibinfo{pages}{473} (\bibinfo{year}{1979}).

\bibitem[{\citenamefont{{Kippenberg} and {Vahala}}(2008)}]{Kippenberg_SCI08}
\bibinfo{author}{\bibfnamefont{T.~J.} \bibnamefont{{Kippenberg}}}
  \bibnamefont{and} \bibinfo{author}{\bibfnamefont{K.~J.}
  \bibnamefont{{Vahala}}}, \bibinfo{journal}{Science}
  \textbf{\bibinfo{volume}{321}}, \bibinfo{pages}{1172} (\bibinfo{year}{2008}).

\bibitem[{\citenamefont{Marquardt and Girvin}(2009)}]{Marquardt_Physics.2.40}
\bibinfo{author}{\bibfnamefont{F.}~\bibnamefont{Marquardt}} \bibnamefont{and}
  \bibinfo{author}{\bibfnamefont{S.~M.} \bibnamefont{Girvin}},
  \bibinfo{journal}{Physics} \textbf{\bibinfo{volume}{2}}, \bibinfo{eid}{40}
  (\bibinfo{year}{2009}).

\bibitem[{\citenamefont{Braginsky and Khalili}(1992)}]{BraginskyKhalili.book}
\bibinfo{author}{\bibfnamefont{V.~B.} \bibnamefont{Braginsky}}
  \bibnamefont{and} \bibinfo{author}{\bibfnamefont{F.~Y.}
  \bibnamefont{Khalili}}, \emph{\bibinfo{title}{{Quantum measurement}}}
  (\bibinfo{publisher}{Cambridge Univ. Press}, \bibinfo{address}{Cambridge},
  \bibinfo{year}{1992}).

\bibitem[{\citenamefont{{Law}}(1995)}]{Law_PRA95}
\bibinfo{author}{\bibfnamefont{C.~K.} \bibnamefont{{Law}}},
  \bibinfo{journal}{Phys. Rev. A} \textbf{\bibinfo{volume}{51}},
  \bibinfo{pages}{2537} (\bibinfo{year}{1995}).

\bibitem[{\citenamefont{{Walls} and {Milburn}}(2008)}]{WallsMilburn.book}
\bibinfo{author}{\bibfnamefont{D.~F.} \bibnamefont{{Walls}}} \bibnamefont{and}
  \bibinfo{author}{\bibfnamefont{G.~J.} \bibnamefont{{Milburn}}},
  \emph{\bibinfo{title}{{Quantum Optics}}} (\bibinfo{publisher}{Springer},
  \bibinfo{year}{2008}).

\bibitem[{\citenamefont{{Scully} and {Zubairy}}(1997)}]{ScullyZubairy.book}
\bibinfo{author}{\bibfnamefont{M.~O.} \bibnamefont{{Scully}}} \bibnamefont{and}
  \bibinfo{author}{\bibfnamefont{M.~S.} \bibnamefont{{Zubairy}}},
  \emph{\bibinfo{title}{{Quantum Optics}}} (\bibinfo{publisher}{Cambridge Univ.
  Press}, \bibinfo{year}{1997}).

\bibitem[{\citenamefont{{Steuernagel} and
  {Scheel}}(2004)}]{Steuernagel_Scheel_JOB04}
\bibinfo{author}{\bibfnamefont{O.}~\bibnamefont{{Steuernagel}}}
  \bibnamefont{and} \bibinfo{author}{\bibfnamefont{S.}~\bibnamefont{{Scheel}}},
  \bibinfo{journal}{J. Opt. B: Quant. Semiclass. Opt.}
  \textbf{\bibinfo{volume}{6}}, \bibinfo{pages}{66} (\bibinfo{year}{2004}),
  \eprint{quant-ph/0211182v1}.

\bibitem[{\citenamefont{{Dorner} et~al.}(2009)\citenamefont{{Dorner},
  {Demkowicz-Dobrzanski}, {Smith}, {Lundeen}, {Wasilewski}, {Banaszek}, and
  {Walmsley}}}]{Dorner_PRL09}
\bibinfo{author}{\bibfnamefont{U.}~\bibnamefont{{Dorner}}},
  \bibinfo{author}{\bibfnamefont{R.}~\bibnamefont{{Demkowicz-Dobrzanski}}},
  \bibinfo{author}{\bibfnamefont{B.~J.} \bibnamefont{{Smith}}},
  \bibinfo{author}{\bibfnamefont{J.~S.} \bibnamefont{{Lundeen}}},
  \bibinfo{author}{\bibfnamefont{W.}~\bibnamefont{{Wasilewski}}},
  \bibinfo{author}{\bibfnamefont{K.}~\bibnamefont{{Banaszek}}},
  \bibnamefont{and} \bibinfo{author}{\bibfnamefont{I.~A.}
  \bibnamefont{{Walmsley}}}, \bibinfo{journal}{Phys. Rev. Lett.}
  \textbf{\bibinfo{volume}{102}}, \bibinfo{pages}{040403}
  (\bibinfo{year}{2009}), \eprint{quant-ph/0807.3659}.

\bibitem[{\citenamefont{{Gerry} and {Knight}}(2005)}]{GerryKnight.book}
\bibinfo{author}{\bibfnamefont{C.~G.} \bibnamefont{{Gerry}}} \bibnamefont{and}
  \bibinfo{author}{\bibfnamefont{P.~L.} \bibnamefont{{Knight}}},
  \emph{\bibinfo{title}{{Introductory Quantum Optics}}}
  (\bibinfo{publisher}{Cambridge Univ. Press}, \bibinfo{year}{2005}).

\end{thebibliography}

\end{document}